\documentclass[aps,pre,twocolumn,showpacs,preprintnumbers,amsmath,amssymb]{revtex4}
\usepackage{graphicx}
\usepackage{dcolumn}
\usepackage{bm}

\begin{document}

\title{Unfolding the procedure of characterizing recorded ultra low frequency, kHZ and MHz electromagnetic anomalies prior to the L'Aquila earthquake as pre-seismic ones. Part II}

\author{K. Eftaxias}
\affiliation{Section of Solid State Physics, Department of Physics, University of Athens, Panepistimiopolis, Zografos, 
15784, Athens, Greece}
\author{G. Balasis}
\affiliation{Institute for Space Applications and Remote Sensing, National Observatory of Athens, Metaxa and Vas. 
Pavlou, Penteli, 15236, Athens, Greece}
\author{Y. Contoyiannis}
\affiliation{Section of Solid State Physics, Department of Physics, University of Athens, Panepistimiopolis, Zografos, 
15784, Athens, Greece}
\author{C. Papadimitriou}
\affiliation{Section of Solid State Physics, Department of Physics, University of Athens, Panepistimiopolis, Zografos, 
15784, Athens, Greece}
\author{M. Kalimeri}
\affiliation{Section of Solid State Physics, Department of Physics, University of Athens, Panepistimiopolis, Zografos, 
15784, Athens, Greece}
\author{J. Kopanas}
\affiliation{Section of Solid State Physics, Department of Physics, University of Athens, Panepistimiopolis, Zografos, 
15784, Athens, Greece}
\author{G. Antonopoulos}
\affiliation{Section of Solid State Physics, Department of Physics, University of Athens, Panepistimiopolis, Zografos, 
15784, Athens, Greece}
\author{C. Nomicos}
\affiliation{Department of Electronics, Technological Educational Institute of Athens, Ag. Spyridonos, Egaleo, 12210, 
Athens, Greece}







\begin{abstract}
Ultra low frequency-ULF (1 Hz or lower), kHz and MHz electromagnetic (EM) anomalies were recorded prior to the 
L'Aquila catastrophic earthquake (EQ) that occurred on April 6, 2009. The detected anomalies followed this 
temporal scheme. (i) The MHZ EM anomalies were detected on March 26, 2009 and April 2, 2009. The kHz EM anomalies 
were emerged on April, 4 2009. The ULF EM anomaly was continuously recorded from March 29, 2009 up to April 2, 
2009. ``Are EQs predictable?'' is a question hotly debated in the science community. Its answer begs for another 
question: ``Are there credible EQ precursors?''. Despite fairly abundant circumstantial evidence pre-seismic EM 
signals have not been adequately accepted as real physical quantities. Therefore, the question effortlessly 
arises as to whether the observed anomalies before the L'Aquila EQ were seismogenic or not. The main goal of this 
work is to provide some insight into this issue. More precisely, the main aims of this contribution are 
threefold: (i) How can we recognize an EM observation as pre-seismic one? We wonder whether necessary and 
sufficient criteria have been established that permit the characterization of an EM observation as a precursor. 
We suggest a procedure for the designation of detected kHz-MHz EM anomalies as seismogenic ones. We do not expect 
to be able to provide a succinct and solid definition of a pre-earthquake kHz-MHz EM emission. Instead, we aim, 
through a multidisciplinary analysis to provide some elements of a definition. (ii) How can we link an individual 
EM precursor with a distinctive stage of the EQ preparation process? The present analysis is consistent with the 
hypothesis that the kHz EM anomalies were associated with the fracture of asperities that were distributed along 
the L'Aquila fault sustaining the system, while the MHz EM anomalies could be triggered by fractures in the 
highly disordered system that surrounded the backbone of asperities of the activated fault. (iii) How can we 
identify precursory symptoms in an individual EM precursor that indicate that the occurrence of the EQ is 
unavoidable? We clearly state that the detection of a MHz EM precursor does not mean that the occurrence of EQ is 
unavoidable; the abrupt emergence of kHz EM emissions indicate the fracture of asperities. The observed ULF EM 
anomaly supports the hypothesis of a relationship between processes produced by increasing tectonic stresses in 
the Earth's crust and attendant EM interactions between the crust and ionosphere. We emphasize that we attempt to 
specify not only whether or not a single EM anomaly is pre-seismic in itself, but mainly  whether a combination 
of emergent ULF, MHz and kHz EM anomalies could be characterized as pre-earthquake.
\end{abstract}

\maketitle

\section{Introduction}

A catastrophic EQ occurred on April 6, 2009 (01h 32m 41s UTC) in central Italy with magnitude Mw = 6.3. The majority of the damage occurred in the city of L'Aquila, the medieval capital city of the Abruzzo region (Fig. 1). 

EM anomalies of a broad frequency range, namely, from ULF (1 Hz or lower), kHz up to MHz (Figs. 2--5), were detected by the sensors located at a mountainous site of Zante island $(37.76^{o}N-20.76^{o}E)$ in the Ionian Sea (western Greece) prior to the L'Aquila EQ (Fig. 1). 

``Are there credible EQ precursors?'' is a question debated in the science community. Herein, based on a multidisciplinary approach, we examine the possible seismogenic origin of the detected EM anomalies. 

An EQ is a sudden mechanical failure in the Earth's crust, which has heterogeneous structures. It is reasonable to expect that its preparatory process has various facets which may be observed before the final catastrophe through seismic, geochemical, hydrological and EM changes (Uyeda et al., 2009). Therefore, the science of EQ prediction should, from the start, be multi-disciplinary (Uyeda et al., 2009). Possible precursory phenomena include changes in seismic velocities, tilt and strain precursors, hydrologic phenomena, chemical emissions and EM signals (Rundle et al. 2003). Herein, we focus on EM precursors. The comprehensive understanding of the EM response to geodynamic processes occurring in the Earth's upper crust is a challenging task in modern geophysics.

Pre-seismic EM emissions have been internationally observed before large EQs (Hayakawa and Fuzinawa, 1994; Hayakawa, 1999; Hayakawa and Molchanov, 2002; Varotsos, 2005; Pulinets and Boyarchuk, 2005; Eftaxias et al., 2007a). These seismo-EM signals may be conveniently classified into the following two major classes (Uyeda et al., 2009 and references therein): 

(i) {\it Direct EM precursors: EM signals believed to be emitted from within the focal zones}. 

EM precursors of a broad frequency range, from low frequency (1 Hz or lower), kHz up to MHz, have been internationally reported before large EQs (Uyeda et al., 2009 and references therein). 

Seismic electric signals (SES), which are low frequency (1 Hz or lower) transient anomalies in telluric current, 
precede EQs, with a lead time from several hours to a few months (Varotsos, 2005). Pressure stimulated 
polarization currents (PSPC), which are emitted from solid containing electric dipoles upon a gradual increase of 
the pressure (or stress) can be a plausible source among other for SES generation (Varotsos and Alexopoulos, 
1986). This emission occurs when the stress reaches a critical value which does not have to coincide with the 
fracture stress. Uyeda et al. (2009) note that, the PSPC model is unique among other models in that SES would be 
generated spontaneously during the gradual increase of stress without requiring any sudden change of stress such 
as micro-fracturing. The electro-kinetic effect, also called {\it streaming potential}, can also be a plausible 
source for SES EM emissions (Uyeda et al., 2009 and references therein). Freund and his colleagues have recently 
been proposing a unique mechanism for ULF electric signals (Freund et al., 2006). They have discovered in 
laboratory that when a block of igneous rock is put under stress locally, the rick turns into a battery, which 
generates its own electric field. 

On the other hand, when a material is strained, intense acoustic and EM emissions, ranging from MHz to kHz, are produced by opening cracks when failure is approached. A stressed rock behaves like a stress-EM transformer. These precursors are detectable both at laboratory and geological scale (Hayakawa and Fujinawa, 1994; Hayakawa, 1999; Gershenzon and Bambakidis, 2001; Hayakawa and Molchanov, 2002; Bahat et al., 2005; Eftaxias et al., 2007a; Muto et al., 2007; Hadjicontis et al., 2007). Studies on the small (laboratory) scale reveal that the MHz EM radiation appears earlier than the kHz one, while the kHz EM emission is launched from 97\% up to 100\% of the corresponding failure strength (Eftaxias et al., 2002 and references therein). On the large (geological) scale, intense MHz and kHz EM emissions precede EQs that: (i) occurred in land (or near coast-line), (ii) were strong (magnitude 6 or larger), (iii) were shallow (Eftaxias et al., 2002, 2004, 2006, 2007b; Kapiris et al., 2004; Karamanos et al., 2006).  Their lead time is ranged from a few days to a few hours. Importantly, the MHz radiation precedes the kHz one at geophysical scale, as well (Eftaxias et al., 2002; Kapiris et al., 2004; Contoyiannis et al., 2005). No co-seismic EM anomalies have been observed (Uyeda et al., 2009 and references therein). 

Notice, a complete sequence of SES, MHz and kHz EM anomalies have been observed one after the other in a series of significant EQs that occurred in Greece (Eftaxias et al., 2000, 2002; Karamanos et al., 2006; Papadimitriou et al., 2008). 

We argue that the detected MHz and kHz EM anomalies on 4/4/2009                                                were emitted from the focal area of the L'Aquila EQ. 

(ii) {\it Indirect EM precursors: EM phenomena believed to be rooted in seismo-ionospheric coupling}

Ample experimental evidence suggests that the preparation of EQ induces a lithosphere-atmosphere-ionosphere (LAI) 
coupling mechanism where EM precursory phenomena are originated (Hayakawa and Fujinawa, 1994; Gokhberg et al., 
1995; Hayakawa, 1999; Hayakawa and Molchanov, 2002; Pulinets et al., 2003; Pulinets and Boyarchuk, 2005; Uyeda et 
al., 2009 and references therein). The regional but substantially {\it large-scale} changes over seismically 
active areas before the seismic shock are mapped into the ionosphere. Both natural and artificial EM signals 
propagate in the Earth-ionosphere waveguide. Any change in the lower ionosphere due to an induced pre-seismic 
LAI-coupling may result in significant changes in the signal propagation-received at a station. We suggest that 
the recorded ULF EM anomaly fits in this category.

Despite fairly abundant circumstantial evidence pre-seismic EM signals have not been adequately accepted as real 
physical quantities. It seems appropriate at this point to admit that there may be legitimate reasons for the 
critical views. The degree to which we can predict a phenomenon is often measured by how well we understand it. 
However, many questions about fracture processes remain standing. {\it Kossobokov (2006) emphasizes that no 
scientific prediction is possible without exact definition of the anticipated phenomenon and the rules, which 
define clearly in advance of it whether the prediction is conformed or not}.  

We pay attention to the fact that the time lags between pre-earthquake EM anomalies and EQs are different for different types of precursors. The remarkable asynchronous appearance of precursors indicates that they refer to different stages of EQ preparation process. Moreover, it implies a different mechanism for their origin. {\it Scientists ought to attempt to link the available various EM observations that appear one after the other to the consecutive processes occurring in Earth's crust}. The comprehensive understanding of EM precursors in terms of physics is an important research target. {\it In our opinion that is a path to achieve more sufficient knowledge of last stages of the EQ preparation process and thus more sufficient short-term EQ prediction. A seismic shift in thinking towards basic science will result a renaissance of strict definitions and systematic experiments in the field of EQ prediction (Cyranoski, 2004)}.

\subsection{Data collection}

Our main tool is the monitoring of the fractures, which occur in the focal area before the final breakup, by recording their MHz-kHz EM emissions. Since 1994, a station has been functioning at a mountainous site of Zante island in the Ionian Sea (western Greece) with the following configuration: (i) six loop antennas recording the three components (EW, NS, and vertical) of the variations of the magnetic field at 3 kHz and 10 kHz respectively; (ii) two vertical $\lambda/2$ electric dipole antennas recording the electric field variations at 41 and 54 MHz respectively. Clear MHz-kHz EM anomalies have been detected over periods ranging from a few days to a few hours before {\it shallows land-based (or near coast-line) EQs with magnitude 6 or larger} since the installation of the station. Recent results indicate that the observed precursors contain characteristic pre-fracture features (Eftaxias et al., 2002, 2004, 2006, 2007b, 2009; Kapiris et al., 2004; Contoyiannis et al., 2005, 2008; Karamanos et al., 2006; Kalimeri et al., 2008; Papadimitriou et al., 2008). 

In order to detect indirect ionospheric precursors we have installed two Short Thin Wire Antennas (STWA) of 100m length each, lying in the Earth's surface in the EW and NS directions, respectively. The analog sensitive device that measures the potential difference in each antenna includes a low-pass active filter with a cut-off frequency of 1 Hz. Thus, the system records ultra low frequency ($<1$ Hz) electric anomalies. Clear ULF EM anomalies have been detected a few days before {\it shallows land-based (or near coast-line) EQs with magnitude 6 or larger} since the installation of the station (Eftaxias et al., 2000, 2002, 2004; Karamanos et al., 2006). All the EM time-series are sampled once per second. 

{\it We stress that the experimental arrangement affords us the possibility to determine not only whether or not a single EM anomaly is pre-seismic in itself, but also whether a combination of ULF, MHz and kHz EM anomalies can be characterized as pre-seismic}.  

The L'Aquila EQ was a very shallow (Walters et al., 2009) land-based event with magnitude 6.3. A sequence of MHz, kHz and ULF EM anomalies was observed before this catastrophic event. The observed anomalies followed this temporal scheme: 

(i) Two MHz EM anomalies appeared on March 26, 2009 and April 2, 2009 (Fig. 2).

(ii) A sequence of strong impulsive kHz EM bursts was emerged on April 4, 2009 (Fig. 3). Fig. 4 shows magnified images of three kHz EM bursts and an excerpt of noise marked in Fig. 3. 

No co-seismic KHz-MHz EM anomalies were observed. 

In this field of research the repeatability of results is desirable. The Chinese (Qian et al., 1994) and Greek (Eftaxias, et al., 2000, 2002; Contoyiannis et al., 2005; Contoyiannis and Eftaxias, 2008) experience can be summarized in the following points: (i) the frequency band of the detected EM anomalies is quite wide; (ii) anomalies are detected earlier in the electric field (MHz) than in the magnetic field (kHz); (iii) the anomalies stop before the EQ occurrence. 

(iii) The ULF anomaly continuously appeared from March 29, 2009 up to April 2, 2009 (Fig. 5). 

Importantly, based on very low frequency (kHz) radio sounding, Rozhnoi et al. (2009) and Biagi et al. (2009) have observed ionospheric perturbations in the time interval 2--8 days before the L'Aquila EQ. 

We note that all the recorded EM anomalies we report here have been obtained during a quiet period in terms of magnetic storms and solar flares activity. 

\subsection{The aim of this work}

We have yet pointed out that despite fairly abundant circumstantial evidence pre-seismic EM signals have not been adequately accepted as real physical quantities (Uyeda et al., 2009). Some open questions in this field of research are:

(i) {\it How can we recognize an EM observation as a pre-seismic one? 
We wonder whether sufficient and necessary criteria have been established that permit the characterization of an EM observation as an EM precursor}. 
 
(ii) {\it How can we link an individual EM precursor with a distinctive stage of the EQ preparation process?}.
 
(iii) {\it How can we identify crucial precursory symptoms in a crucial EM observation that indicate that the occurrence of the EQ is unavoidable?}.

Here we shall study the possible seismogenic origin of EM anomalies recorded prior to the L'Aquila EQ within the frame work of the above questions. More precisely:

(i) {\it A main aim of this contribution is to suggest a procedure for the designation of the emergent EM anomalies as seismogenic ones. We do not expect to be possible to provide a succinct and solid definition of pre-seismic EM emissions. We attempt, through a multidisciplinary analysis, to provide elements of a definition}.
 
(ii) {\it We try to link the detected MHz and kHz EM anomalies with two equivalent stages in the L'Aquila EQ preparation process}.

(iii){\it We make an effort to put forward physically meaningful arguments to support a way of quantifying the time to global failure and the identification of launch of a key EM precursor beyond which the evolution towards global failure becomes irreversible}. 

\section{A two stages model of EQ preparation in terms of MHz-kHz EM emission}

An important challenge in this field of research is to distinguish characteristic epochs in the evolution of the EM response to geodynamic processes occurring in the Earth's upper crust and identify them with the equivalent stages in the EQ preparation process. {\it Any result of such an approach, either a success or a failure, clarifies our knowledge about seismic processes supporting or rejecting the underlying conceptions}. 

We recall that an important feature, observed both on laboratory and geophysical scale, is that the MHz radiation precedes the kHz one. Recently, we proposed that the pre-earthquake MHz and kHz EM anomalies obey the following two-stage model (Kapiris et al., 2004; Contoyiannis et al., 2005, 2008; Papadimitriou et al., 2008; Eftaxias et al., 2006, 2007b): (i) The MHz EM emission is thought to be due to the fracture of the highly heterogeneous system that surrounds the family of large high-strength asperities distributed along the fault sustaining the system. (ii) The kHz EM radiation is due to the fracture of the asperities themselves. By the end of this paper we will argue that the emergent MHz and kHz EM anomalies prior to the L'Aquila EQ also follow the aforementioned two stages model. 

\section{Focus on the possible seismogenic origin of the observed MHz EM anomalies}  

Fracture process in heterogeneous materials is characterized by fundamental properties. Especially, such a fracture process: (i) is characterized by anti-persistency; and (ii) can be described in analogy with a thermal continuous second order phase transition. These two crucial footprints should be mirrored on the associated MHZ EM precursor (Kapiris et al., 2004, Contoyiannis et al., 2005). 

In natural rocks at large length scales there are long-range $\it anti-correlations$, in the sense that a high value of a rock property, e.g. threshold for breaking, is followed by a low value and vice versa. Failure nucleation begins to occur at a region where the resistance to rupture growth has the minimum value. An EM event is emitted during this fracture. The fracture process continues in the same weak region until a much stronger region is encountered in its neighbourhood. When this happens, fracture stops, and thus the emitted EM emission sieges.  The stresses are redistributed, while the applied stress in the focal area increases. A new population of cracks nucleates in the weaker of the unbroken regions, and thus a new EM event appears, and so on. In summary, the interplay between the heterogeneities and the stress field in the focal area should be responsible for the appearance of $\it antipersistency$ in a pre-seismic MHz EM time series (Kapiris et al., 2004; Contoyiannis et al., 2005; Contoyiannis and Eftaxias, 2008). In the following, we will show that the detected MHz activity before the L'Aquila EQ is really characterized by anti-persistency. 

Physically, the presence of anti-persistency implies a set of EM fluctuations tending to induce {\it stability} to the system, essentially the existence of a {\it non-linear negative feedback mechanism that ``kicks'' the opening rate of cracks away from extremes}. The existence of such a mechanism leads to the next step: {\it in analogy to the study of critical phase transitions in statistical physics, it has been proposed that the fracture of heterogeneous materials can be described in analogy with a thermal continuous second order phase transition in equilibrium (Herrmann and Roux, 1990; Sornette, 2004)}. We will show that the observed MHz EM anomalies can be described such as critical phenomena by means of the recently introduced Method of Critical Fluctuations (MCF) (Contoyiannis and Diakonos, 2000; Contoyiannis et al., 2002). We note that MHz EM anomalies detected prior to significant EQs in Greece follows the aforementioned critical behaviour (Contoyiannis et al., 2005, 2009).

\subsection{The method of critical fluctuations}

The MCF, which constitutes a statistical method of analysis for the critical fluctuations in systems that undergo a continuous phase transition at equilibrium, has been recently introduced (Contoyiannis and Diakonos 2000; Contoyiannis et al., 2002). The authors have shown that the fluctuations of the order parameter $\phi$, that correspond to successive configurations of critical systems at equilibrium, obey a dynamical law of intermittency which can be described in terms of a 1-d nonlinear map. The invariant density $\rho(\phi)$ for such a map is characterized by a plateau which decays in a super-exponential way [see Fig. 1 in (Contoyiannis and Diakonos, 2000)]. The exact dynamics at the critical point can be determined analytically for a large class of critical systems introducing the so-called critical map (Contoyiannis and Diakonos, 2000). For small values of $\phi $, this critical map can be approximated as  

\begin{equation}
\phi_{n+1}=\phi_n + u \phi_n^z +\epsilon_n
\end{equation}

The shift parameter $\epsilon_n$ introduces a non-universal stochastic noise: each physical system has its characteristic ``noise'', which is expressed through the shift parameter $\epsilon_n$. For thermal systems the exponent $z$ is introduced, which is related to the isothermal critical exponent $\delta$ by $z= \delta + 1$.  

The plateau region of the invariant density $\rho(\phi)$ corresponds to the laminar region of the critical map where fully correlated dynamics takes place (Contoyiannis et al., 2005 and references therein). The laminar region ends when the second term in Eq. (1) becomes relevant. However, due to the fact that the dynamical law (1) changes continuously with $\phi $, the end of the laminar region cannot be easily defined based on a strictly quantitative criterion. Thus, the end of the laminar region should be generally treated as a variable parameter.

Based on the foregoing description of the critical fluctuations, the MCF develops an algorithm permitting the extraction of the critical fluctuations, if any, in a recorded time series. The important observation in this approach is the fact that the distribution $ P(l) $ of the laminar lengths $l$ of the intermittent map (1) in the limit  $\epsilon_n \to 0 $ is given by the power law (Contoyiannis et al., 2002)  

\begin{equation}
P(l) \sim l^{-p_l}
\end{equation}

where the exponent $p_l$ is connected with the exponent $z$ via $p_l=\frac{z}{z-1}$. Therefore the exponent $p_l$ is related to the isothermal exponent $\delta$ by

\begin{equation}
p_l= 1 + \frac{1}{\delta}
\end{equation}
where   $\delta  >  0$. 

Inversely, the existence of a power law such as relation (2), accompanied by a plateau form of the corresponding density $\rho(\phi)$, is a signature of underlying correlated dynamics similar to critical behavior (Contoyiannis and Diakonos, 2000b; Contoyiannis et al., 2002).
  
We emphasize that it is possible in the framework of universality, which is characteristic of critical phenomena, to give meaning to the exponent $p_l$ beyond the thermal phase transitions (Contoyiannis et al., 2002). The MCF is directly applied to time series or to segments of time series which appear to have a cumulative stationary behavior. 

The main aim of the MCF is to estimate the exponent $p_l$. The distribution of the laminar lengths, $l$, of fluctuations included in a stationary window is fitted by the relation:

\begin{equation}
P(l) \sim  l^{-p_2} e^{-p_3 l}
\end{equation}

If $p_3$ is zero, then $p_2$ is equal to $p_l$.  Practically, as $p_3$ approaches zero, then $p_2$ approaches $p_l$ and the laminar lengths tend to follow a power-law type distribution. So, we expect a good fit to Eq. (4) with $p_2 > 1$ and $p_3 \approx 0$ if the system is in a critical state.

We stress that when the exponent $p_2$ is smaller than one, then, independently of the $p_3$-value, the system is not in a critical state. Generally, the exponents $p_2, p_3$ have a competitive character, namely, when the exponent $p_2$ decreases the associated exponent $p_3$ increases (they are mirror images of each other). In this way, we can identify the deviation from the critical state. To be more precise, as the exponent $p_2$ ($p_2<1$) is close to 1 and simultaneously the exponent $p_3$ is close to zero, then the system is in a sub-critical state. As the system moves away from the critical state, then the exponent $p_2$ decreases while simultaneously $p_3$ increases, reinforcing in this way the exponential character of the laminar length distribution. 

In summary, the research of criticality in natural systems could be quantitatively accomplished by estimating the values of only two parameters, namely the exponents $p_2$ and $p_3$. 

Up to now, the MCF has been applied on: (i) numerical experiments of thermal systems (Ising models) (Contoyiannis et al., 2002), (ii) EM pre-seismic signals (Contoyiannis et al., 2005), and (iii) electro-cardiac signals from biological tissues (Contoyiannis et al., 2004).

\subsection{Application of the MCF method to the detected pre-seismic MHz EM radiation prior to the L'Aquila EQ}

As it was said, two critical MHz EM anomalies were detected on 26/3/2009 and 2/4/2009 prior to the EQ. In (Contoyiannis et al. 2009) we have shown that the anomaly recorded on 2/4/2009: (i) can be described as analogous to a thermal continuous phase transition, and (ii) has anti-persistent behaviour. These features imply that this candidate precursor could be triggered by fractures in the highly disordered system that surrounded the backbone of asperities of the activated fault (Contoyiannis et al., 2005). 

Herein, we study the MHz anomaly that emerged on 26 March, 2009 (Fig. 2).  The critical time interval includes approximately 10000 points (Fig. 2). The stationary behaviour of the critical window has been checked by estimating the mean value and the standard deviation for various time intervals in the CW, all having a common origin (Contoyiannis et al., 2005).  

Notice, we have shown that the amplitude of the recorded electric time series, or equivalently the output of the detector (in mV), behaves as {\it order parameter} (Contoyiannis et al., 2005). The symbol $\psi $ symbolizes the order parameter in this particular case under study.

The next step is to produce the distribution of the laminar lengths, $l$. The laminar length gives the stay time within the laminar region. More precisely, the laminar lengths are described by the lengths of the sub-sequences in the time series that result from successive $\psi$-values obeying the condition $\psi_o \leq \psi  \leq  \psi_l$, where $\psi_o $ is the fixed point and the end of the laminar region $\psi_l$ is a variable parameter. 

Using the fitting relation (4) we estimate the exponents $p_2, p_3$ for different $\psi_l$- values. In Fig. 6 the exponents $p_2, p_3$ plotted against the end point $\psi_l$ are shown. 

Fig. 6a reveals that the majority of the $p_2$ values are greater than 1 and the corresponding $p_3$-values close to zero for all end points of laminar regions. This finding indicates that all the trajectories are critical and carry essentially the same information about the dynamical term. In terms of physics this means that the system is characterized by a ``{\it strong criticality}''.
Fig. 6b shows an example of the distribution of the laminar lengths for a given end point. The function (Eg. 4) fits the data with $p_2 = 1.48$ and $p_3 = 0.009$, while the quality of fitting is excellent ($r^2 = 0.998$). 

Fig. 7 shows the corresponding critical behaviour of the MHZ EM fluctuations which appeared on 2/4/2009 (Fig. 2). This group of EM fluctuations also shows a ``{\it strong criticality}''. The associated analysis by means of MCF has been presented in details in (Contoyiannis et al., 2009). 

\subsection{The anti-persistent behaviour of the candidate EM precursor}

The exponent $H$ characterizes the persistent/anti-persistent properties of the signal (Hurst, 1954). The range $0<H<0.5$ indicates an anti-persistency, so that increases in the fluctuations within a time internal are likely to be followed by decreases in the following time internal, and conversely. On the contrary, the range $0.5<H<1$ $(2 < \beta <3)$ indicates persistent behavior. This means that increases in the fluctuations within a time internal are likely to be followed by increases in the next internal, so the system starts to govern by a positive feedback process. 

Recently, we introduced the following connection between the exponents $H$ and $p_2$ (Contoyiannis et al., 2005):  

\begin{equation}
H=2-\frac{3}{2} p_l
\end{equation}

As we see from this equation, the allowed range of $p_l$ values for a second-order phase transition, i.e., $p_l >1 $, leads to the condition $ H<0.5 $ which indicates the existence of an anti-persistent mechanism. It was found that Eq. (5) is valid only if the exponent $p_2$ lies in the interval $1<p_2<1.5$ (Contoyiannis et al., 2005). Figs. (6) and (8) show that the $p_2$ values associated with the emergent two MHz EM precursors on March 26, 2009 and April 2, 2009 obey the restriction $1<p_2<1.5$, and thus show anti-persistency.  As it was mentioned, the interplay between the heterogeneities in the pre-focal area and the stress field could be responsible for the observed anti-persistent pattern. 

The question naturally arises as to whether the appearance of criticality and anti-persistency is systematically observed.  Recently, we have listed MHz EM activities recorded prior to nine significant EQs that have been occurred in-land (or near coast line) (Contoyiannis et al., 2009). All these precursors show anti-persistency and can be described in analogy with a thermal continuous second order transition. 

A key-question concerns the physical mechanism that drives the heterogeneous system to its critical state. Combining the ideas of Levy statistics, non-extensive Tsallis statistics (Tsallis, 1988, 2009) and criticality on the one hand, and features included in the precursory MHz time series on the other, we argued that {\it a Levy-walk-type mechanism can drive the heterogeneous system to criticality (Contoyiannis et al., 2008)}. 

We conclude that two fundamental features of fracture of heterogeneous materials have been projected on the detected MHz EM activities. This finding supports the hypothesis that they are associated with the fracture process in the heterogeneous regime of the focal area. 

\subsection{Is the evolution towards global failure irreversible after the appearance of the MHz EM precursor?}

A fundamental question that we ought to address is as follows. {\it Is the evolution towards global failure irreversible after the appearance of the MHz EM precursor?} We clearly state in (Contoyiannis et al., 2005) that the detection of a MHz EM precursor, which could be described in analogy with a thermal continuous second order transition and shows anti-persistent behaviour, does not mean that the occurrence of EQ is unavoidable. Its appearance reveals that the fracture of heterogeneous system in the focal area has been obstructed along the backbone of asperities that sustain the system. {\it The ``siege'' of strong asperities begins. The EQ will occur if and when the local stress exceeds the fracture stress of asperities}. We argue that the abrupt emergence of kHz EM emissions indicates the fracture of asperities and thus signalizes that the evolution of the process toward global failure is unavoidable (Kapiris et al., 2004; Contoyiannis et al., 2005; Papadimitriou et al., 2008, Eftaxias, 2009). In our opinion, the appearance of a critical anti-persistent MHz EM anomaly is a {\it necessary} but not {\it sufficient} condition for the EQ occurrence. We have detected such MHz EM anomalies which, however, were not accompanied by a significant EQ. We emphasize that: (i) following the procedure, which has been reported in (Koulouras et al, 2009), we have excluded the possibility that these anomalies were related to magnetic storm activity or solar flare activity; (ii) the emergence of these MHz anomalies were not accompanied by the appearance of kHz and ULF EM anomalies.  It is an open question whether these critical and anti-persistent ``strange'' anomalies were seismogenic ones or not. 

\section{Focus on the possible seismogenic origin of the detected kHz EM anomaly}  

In Part I of this communication (Eftaxias et al., 2009b) we focus on the detected kHz EM anomaly, which plays a crucial role in our approach. We try to introduce criteria in order to discriminate this anomaly from background noise. For this purpose, we analyze the data successively in terms of various concepts of entropy and information theory, namely, Shannon $n$-block entropy, conditional entropy, entropy of the source, Kolmogorov-Sinai entropy, $T$-entropy, approximate entropy, fractal spectral analysis, R/S analysis and detrended fluctuations analysis. All the techniques suggest that the kHz EM anomaly is characterized by: (i) a significant lower complexity (or higher organization, lower uncertainty, higher predictability, higher compressibility); and (ii) strong persistency. The simultaneous appearance of both these two characteristics implies that the underlying fracto-electromagnetic process is governed by a {\it positive feedback mechanism}. Such a mechanism is consistent with the anomaly's being a precursor of an ensuing catastrophic event. Notice, the aforementioned two crucial signatures are also hidden in other quiet different complex catastrophic events (epileptic seizures, magnetic storms, solar flares) as predicted by the theory of complex systems (Eftaxias et al., 2006; Balasis et al., 2006; Koulouras et al., 2009). 

In summary, the application of the procedure outlined in Part I (Eftaxias et al., 2009b) clearly recognizes and discriminates the candidate kHz EM precursors from the EM background in the region of the station. However, by the end of Part I (Eftaxias et al., 2009b) we conclude that the results of this multidisciplinary analysis by themselves are not sufficient to characterize the kHz EM anomalies as pre-earthquake. They offer {\it necessary} but not {\it sufficient} criteria in order to recognize an emergent kHz EM anomaly as a fracto-electromagnetic one. Much remains to be done to tackle systematically real pre-seismic EM precursors. 

\subsection{Our strategy}  

The Earth's crust is extremely complex. However, despite its complexity, there are several universally holding scaling relations. Such universal structural patterns of fracture and faulting process should be included into an associated fracto-electromagnetic activity. Therefore, an important pursuit is to make a quantitative comparison between {\it temporal fractal patterns} of the precursory kHz EM time series on one hand and {\it universal spatial fractal patterns} of fracture surfaces on the other hand. Notice, Maslov et al. (1994) have formally established the relationship between spatial fractal behavior and long-range temporal correlations for a broad range of critical phenomena. They showed that {\it both the temporal and spatial activity can be described as different cuts in the same underlying fractal}. A self-organized critical process, as the source of the temporal power-laws, would further suggest that similar power-laws exist also for parameters in the spatial domain (Hansen and Schmittbuhl, 2003). A time series of major historical events could have temporal and spatial correlations. Laboratory experiments support the consideration that both the temporal and spatial activity can be described as different cuts in the same underlying fractal. Characteristically, Ponomarev et al. (1997) have studied in the laboratory the temporal evolution of Hurst exponent for the series of distances $H_r$ and time intervals $H_t$ between consecutive acoustic emission events in rocks. Their analysis indicates that the changes of $H_r$ and $H_t$ with time occur in phase, while the relationship $H_r \approx H_t$ is valid. 

The study of the morphology of fracture surfaces is nowadays a very active field of research. From the early work of Mandelbrot (1982), much effort has been put into the statistical characterization of the resulting fractal surfaces in fracture processes: (i) Fracture surfaces were found to be self-affine following the fBm model over a wide range of length scales. (ii) The spatial roughness of fracture surfaces has been interpreted as a universal indicator of surface fracture, weakly dependent on the nature of the material and on the failure mode. 
{\it Therefore, two questions naturally arise as to whether the kHz EM activity detected prior to the L'Aquila EQ: (i) constitutes a temporal fractal following the fBm model; and (ii) has a temporal profile with roughness which is in harmony with the universal spatial roughness of fracture surfaces.} 

Moreover, the aspects of self-affine nature of faulting and fracture have been well-established. {\it Consequently, two additional questions arise as to whether the kHz EM precursor is: (i) a reduced image of the regional seismicity; and (ii) a magnified image of laboratory seismicity.} In the subsequent subsections we attempt to answer the above questions.

\subsection{Signatures of fractional-Brownian-motion nature of faulting and fracture in the candidate kHz EM precursor}

The aspect of self-affine nature of faulting and fracture is widely documented. Especially, ample experimental and theoretical evidence supports the existence of a fractional Brownian motion (fBm) scheme: (i) Kinematic or dynamic source inversions of EQs suggest that the final slip (or the stress drop) has a heterogeneous spatial distribution over the fault [see among others Gusev, (1992); Bouchon (1997); Peyrat et al., (2001)]. (ii) Power spectrum analysis of the fault surface suggests that heterogeneities are observed over a large range of scale lengths [see Power et al., (1987), in particular Fig. 4]. (iii) Investigators of the EQ dynamics have already pointed out that the fracture mechanics of the stressed crust of the Earth forms self-similar fault patterns, with well-defined fractal dimensionalities (Kagan et al., 1982; Sahimi et al., 1993; Barriere and Turcotte, 1991). (iv) Following the observations of the self-similarity in various length scales in the roughness of the fractured solid surfaces, Chakrabarti et al. (1999) have proposed that the contact area distribution between two fractal surfaces follows a unique power law. (vi) Huang and Turcotte (1988) have pointed out that natural rock surfaces can be represented by fBm over a wide range. 

If a time series is a temporal fractal then a power-law of the form $S(f) \propto f^{-\beta}$ is obeyed, with $S(f)$ the power spectral density and $f$ the frequency. In Part I of this contribution (Eftaxias et al., 2009b) we shown that the emergent strong kHz EM fluctuations on April 4, 2009 follows the law $S(f) \propto f^{-\beta}$, while the $\beta$ exponent takes high values, i.e., between 2 and 3. This finding indicates that the temporal profile of the observed EM bursts actually follows the fBm-model (Heneghan and McDarby, 2000). We review this analysis in Fig. 8. {\it In summary, the universal fBm profile of the fracture surfaces has been mirrored in the temporal profile of the kHz EM activity under study}.   

\subsection{Footprints of universal roughness value of fracture surfaces in the kHz EM activity}

As it was said, fracture surfaces are self-affine following the fBm-model over a wide range of length scales. The height-height correlation function
 
$\Delta h(ƒr) = <[h(r +\Delta r) - h(r)]_r ^{1/2}$ 

computed along a given direction is found to scale as $\Delta h \sim (\Delta r)^H$ , where $H$ refers to the Hurst exponent. $H$ specifies the strength of the irregularity (``roughness'') of the fBm surface topography: the fractal dimension is calculated from the relation $D = (2 - H)$ (Heneghan and McDarby, 2000). The ``roughness'' exponent $H$ expresses the tendency for $dh = [dh(x)/dx]dx$ to change sign. When $1/2 < H < 1$, the sign tends not to change. The value $H = 1$ is an upper bound reached when the ``roughness'' of the fault is minimum, in other words, a differentiable surface topography corresponds to $H = 1$. When $H = 1/2$, the sign of $dh$ changes randomly, and the corresponding surface possesses no spatial correlations. For $0 < H < 1/2$ there is a tendency for the sign to change (anticorrelation). The value $H = 0$ is a lower bound; as $H$ tends towards 0 trends are more rapidly reversed giving a very irregular look. 

We paid attention to the fact that {\it the Hurst exponent $H \sim 0.7-0.8$ has been interpreted as a {\it universal indicator} of surface fracture, weakly dependent on the nature of the material and on the failure mode} (Lopez and Schmittbuhl., 1998; Hansen and Schmittbuhl., 2003; Ponson et al., 2006; Mourot et al., 2006; Zapperi et al., 2005). 

In Part I of this work (Eftaxias et al., 2009b) we showed that the ``roughness'' of the profile of the kHz EM time series, as it represented by the associated Hurst exponent, is distributed around the value 0.7. This result has been verified by means of both fractal spectral analysis and R/S analysis.

Importantly, the surface roughness of a recently exhumed strike-slip fault plane has been measured by three independent 3D portable laser scanners (Renard et al., 2006). This fault plane refers to the Vuache fault, near Annecy in the French Alps. Statistical scaling analyses show that the striated fault surface exhibits self-affine scaling invariance that can be described by a scaling roughness exponent, $H_1 = 0.7$ in the direction of slip. Fig. 9 shows the $log$-$log$ representation of $S(f) \propto f^{-\beta}$ into a characteristic excerpt of the recorded kHz EM anomaly, namely the EM burst B1. The associated $\beta$ exponent is equal to 2.39, which leads to $H = 0.7$. We recall that the $\beta$ exponent is related to the Hurst exponent, $H$, by the formula $\beta = 2H +1$, for the fBm (Heneghan and McDarby, 2000).

{\it We conclude that the universal spatial roughness of fracture surfaces nicely coincides with the roughness of the temporal profile of the kHz EM anomaly that was emerged on April 4, 2009, i.e., a few tens of hours prior to the L'Aquila EQ}. 

The observed low ``roughness'' in the recorded kHz EM bursts is consistent with the following physical picture. As the two rough surfaces of the activated fault relatively move, the "details" in the topography of two surfaces are gradually removed, thus there is an increase in the population of larger asperities with time with respect to that of small asperities. Simultaneously, contacts become more closely spaced. In this way a high Hurst exponent decreases the number but increases the size of contacts / kHz EM events. In summary, the observed more or less regularity in the temporal profile of the recorded kHz EM anomaly is in harmony with the fracture of a family of large, strong and closely spaced asperities. 

A series of systematic high-resolution laboratory experiments have been performed by Ohnaka and Shen (1999) on the nucleation of propagating slip failure on pre-existing faults having different surface roughness to demonstrate how the size scale and duration on shear rupture nucleation are affected by geometric irregularity of the rupturing surfaces. The authors conclude that the rougher the rupturing surfaces, the greater the timescales of rupture nucleation are. In the frame of this study, the observed short duration of the recorded kHz EM bursts is consistent with the estimated low ``roughness''.  

\section{The activation of the L'Aquila fault as a reduced self-affine image of regional natural seismicity and a magnified image of laboratory seismicity}

The aspect of self-affine nature of faulting and fracture is widely documented from both, field observations and laboratory experiments, and studies of failure precursors on the small (laboratory) and large (EQ) scale (Mandelbrot, 1982; Huang and Turcotte, 1988; Turcotte, 1992; Rabinivitch et al., 2001; Rundle et al., 2003; Sornette, 2004; Muto et al., 2007). This fundamental aspect bridges the activation of a single fault with the regional seismicity on one hand and laboratory seismicity on the other hand. In the following we examine whether the kHz EM precursor verifies this prospect. More precisely, {\it we try to show that the activation of the L'Aquila fault is a reduced self-affine image of the regional seismicity and a magnified image of laboratory seismicity}. Importantly, {\it Huang and Turcotte (1988) have suggested that the statistics of regional seismicity could be merely a macroscopic reflection of the physical processes in EQ source}. For the aforementioned purpose, we are based on: (i) A model for EQ dynamics coming from a non-extensive Tsallis formulation. This model has been proposed by Solotongo-Costa and Posadas (2004) and revised by Silva et al (2006).  (ii) The Gutenberg-Richter (G-R) magnitude-frequency relationship for EQs.

\subsection{The L'Aquila fault activation as a reduced self-affine image of the regional seismicity}

The EQ dynamics model proposed by Solotongo-Costa and Posadas (2004) is starting from first principles. The authors assume that: (i) The mechanism of relative displacement of fault plates is the main cause of EQs. (ii) The space between fault planes is filled with the residues of the breakage of the tectonic plates, from where the faults have originated. The motion of the fault planes can be hindered not only by the overlapping of two irregularities of the profiles, but also by the eventual relative position of several fragments. Thus, the mechanism of triggering EQs is established through the combination of the irregularities of the fault planes on one hand and the fragments between them on the other hand.  The fragments size distribution function comes from a nonextensive Tsallis formulation, starting from first principles, i.e., a nonextensive formulation of the maximum entropy principle. This approach leads to a G-R type law for the magnitude distribution of EQs:

\begin{eqnarray}
\log(N(m>))=\log N+\left(\frac{2-q}{1-q}\right)\times\\\nonumber
\times\log\left[1+\alpha(q-1)\times(2-q)^{(1-q)/(q-2)}
10^{2m}\right]
\end{eqnarray}

where $N$ is the total number of EQs, $N(m>)$ the number of EQs with magnitude larger than $m$, and $m\approx \log (\varepsilon)$. $\alpha$ is the constant of proportionality between the EQ energy, $\varepsilon$ and the size of fragment, $r$. More precisely, Solotongo-Costa and Posadas assume that $\varepsilon \propto r$. We clarify that the entropic index $q$ describes the deviation of Tsallis entropy from the traditional Shannon one.

Silva et al. (2006) have subsequently revised this model considering the current definition of the mean value, i.e., the so-called $q$-expectation value. They also suggested a Gutenberg-Richter type law, which provides an excellent fit to seismicities, too:

\begin{equation}
\log(N_{>m})=\log N+\left(\frac{2-q}{1-q}\right)
\log\left[1-\left(\frac{1-q}{2-q}\right)\left(\frac{10^{2m}}{\alpha^{2/3}}\right)\right]
\end{equation}

Herein $\alpha$ is the constant of proportionality between the EQ energy, $\varepsilon$, and the size of fragment, $r$. 

We emphasize that the proposed non-extensive G-R type laws (6) and (7) provide an excellent fit to seismicities generated in various large geographic areas usually identified as ``{seismic regions'', each of them covering many geological faults: {\it the $q$-values are restricted in the narrow region from 1.60 to 1.71} (Solotongo-Costa and Posadas (2004); Silva et al., 2006; Vilar et al., 2007). Notice, the magnitude-frequency relationships for EQs (6) and (7) do not say anything about a specific activated fault (EQ). Interestingly, Hallgass et al. (1997) have emphasized that {\it what is lacking is the description of what happened locally, i.e., as a consequence of a single event}. The candidate kHz EM precursor refers to a specific (L'Aquila) fault.

When a material is strained, EM emissions are produced by opening cracks. Thus, in the frame of the above-mentioned nonextensive models for EQ dynamics, a precursory sequence of EM bursts occurs when there is fracture of the fragments that fill the space between the irregular fault planes of the activated individual fault. \textit{In the following, we study whether the statistics of regional seismicity could be merely a macroscopic self-affine reflection of the L'Aquila EQ, as it has been early suggested by Huang and Turcotte (1988)}. Recently, we have examined this issue in terms of Solotongo-Costa and Posadas approach (Eftaxias, 2009). Herein, we investigate the universality by means of the revised model for EQ dynamics which has been introduced by Silva et al. (2006) and quantitatively described by Eg. (7). 

We regard as amplitude $A$ of a candidate ``fracto-EM fluctuation'' the difference $A_{fem}(t_{i})=A(t_{i})-A_{noise}$, where $A_{noise}$ is the background (noise)  level of the EM time series. We consider that a sequence of $k$ successively emerged ``fracto-EM fluctuations'' $A_{fem}(t_{i})$, $i=1,\ldots,k$ represents the EM energy released, $\varepsilon$, during the damage of a fragment. We shall refer to this as an ``EM EQ''. Since the squared amplitude of the fracto-EM emissions is proportional to their energy, the magnitude $m$ of the candidate ``EM EQ'' is given by the relation $m=\log\varepsilon \sim \log
\left(\sum\left [ A_{fem}(t_{i})\right]^{2}\right)$.

Fig. (10) shows that Eq. (7) provides an excellent fit to the pre-seismic kHz EM experimental data, incorporating the characteristics of nonextensivity statistics into the distribution of the detected precursory  ``EM EQs'' on 4/4/2009. Herein, $N$ is the total number of the detected ``EM, $N(m>)$ the number of  `EM EQs'' with magnitude larger than $m$, $P(>m)=N(m>)/N$ the relative cumulative number of ``EM EQs'' with magnitude larger than $m$, and $\alpha$ the constant of proportionality between the EM energy released and the size of fragment. {\it The best-fit parameter for this analysis is given by $q=1.82$}. 

Fig. (11) shows that Eq. (7) also provides an excellent fit to the ``EM EQs'' included in the EM burst B1 (see Figs. 3, 4) with {\it $q=1.87$}. 

{\it It is very interesting to observe the similarity in the $q$-values associated with the non-extensive Eq. 7 for: (i) seismicities generated in various large geographic areas, and (ii) the precursory sequence of ``EM EQs'' possibly associated with the activation of the L'Aquila fault}. This finding indicates that the statistics of regional seismicity could be merely a macroscopic reflection of the physical processes in the EQ source. 

{\it Remark}: A very clear kHz EM precursor was emerged prior to the Athens (Greece) EQ (M = 5.9) that occurred on September 7, 1999. Importantly, this precursor also follows Eq. (7) with $q=1.80$ which is nicely close to that associated with the precursor of the L'Aquila EQ ($q=1.82$).

It is worth mentioning that the estimated non-extensive $q$ parameter is in full agreement with the upper limit ${q<2}$ obtained from several studies involving the Tsallis non-extensive framework (Carvalho et al., 2008; Zunino et al., 2008). Moreover, it is in harmony with an underlying sub-extensive system, $q>1$, verifying the emergence of strong interactions in the Earth's crust during the EQ preparation process.

\subsection{The activation of the L'Aquila fault as a magnified self-affine image of the laboratory seismicity} 

It would be helpful to have analyses of laboratory seismicities by means of nonextensive formulae (6) and (7), and thus to compare the corresponding $q$-values with the ones associated with the kHz EM activity under study. Unfortunately, this information is lacking. However, laboratory seismicities have been investigated in terms of the traditional Gutenberg-Richter (G-R) law, which is the best known scaling relation for EQs: the cumulative number of EQ with magnitude greater than $M$ is given by

\begin{equation}
\log N(>m)=\alpha-bm,
\end{equation}

where $N(m>)$ is the cumulative number of EQs with a magnitude greater than m occurring in a specified area and time and b and $\alpha$ are constants. Due to the above mentioned reasons, we examine whether the traditional G-R law also describes the energy distribution of the kHz EM anomalies prior to the L'Aquila EQ.

Following laboratory analyses, in Fig. (12) we depict the quantity $N(>m)$ vs ``fracto-EM fluctuation'' magnitude, $m$, where $N(>m)$ is the cumulative number of ``fracto-EM fluctuations'' with magnitude greater than $m$. The main part of this distribution is given by $\log N(>M)=\alpha-bm$, where $b = 0.52$.
 
We focus on the estimated value $b\sim 0.52$. There are increasing reports on premonitory decrease of $b$-value {\it immediately before} the global fracture. Characteristically, Lei and Satoh (2007), based on acoustic emission events recorded during the {\it catastrophic fracture} of typical rock samples under differential compression, suggest that the pre-failure damage evolution is characterized by a dramatic decrease in $b$-value from $\sim 1.5$ to $\sim 0.5$ for hard rocks. Laboratory experiments in terms of acoustic emission performed by Ponomarev et al. (1997) also showed a significant fall of the observed b-values from $\sim 1$ to $\sim 0.6$ {\it just before} the global rupture. Rabinovitch et al., (2001) found that laboratory piezo-stimulated EM emission follows the G-R law with $b = 0.62$. Notice, the sequence of kHz EM fluctuations recorded prior to the Athens EQ obey the G-R distribution with $b = 0.62$.

The above mentioned results imply that {\it the activation of the L'Aquila fault behaves as a magnified self-affine image of the laboratory seismicity}.

It is known that a significant decrease of $b$-value takes place before EQs, as well: foreshock sequences and main shocks are characterized by a much smaller exponent compared to aftershocks, $b\sim1$. This evidence further verifies that the activation of the L'Aquila fault behaves as a reduced self-affine image of the regional natural seismicity.

The $b$-value represents a statistical measurement of the relative abundance of large and small EQs in the group. The estimated low $b$-value means that a high fraction of the total observed ``EM-EQs'' prior to the L'Aquila EQ occurs at higher magnitudes. This finding is in consistency with the estimated high and low values of $\beta$ and $H$ exponents, respectively. 

\subsection{Evidence of fractional-Brownian-motion-type asperity model for EQ generation in candidate the kHz EM emission associated with the L'Aquila 
EQ}

De Rubeis et al., (1996) and Hallgass et al., (1997) have introduced {\it a self-affine asperity model for the seismicity that mimics the fault friction by means of two fractional Brownian profiles that slide one over the other}. An EQ occurs when there is an overlap of the two profiles representing the two fault faces and its energy is assumed proportional to the overlap surface. This model exhibits a good interpretation by means of the Gutenberg-Richter law of the seismicity generated in {\it a large geographic area} usually identified as ``seismic region'', covering many geological faults, on a global sense.

Hallgass et al., (1997) have stated that {\it ``what is lacking is the description of what happened locally, i.e., as a consequence of a single event, from both the temporal and the spatial point of view''}. In this section, we focus on the activation of a {\it single}, namely L'Aquila fault. In the frame of the proposed self-affine asperity model for EQ dynamics the fracture of an intersection between the two fractional Brownian profiles of a single fault is accompanied by the launch of an individual kHz EM burst ("EM-EQ"). {\textit Therefore, a vital question is whether the sequence of the observed kHz EM pulses under study could be induced by the slipping of two rough and rigid fBm profiles one over the other, or, equivalently, whether the EM activity behaves as a temporal fractal following the fBm model}. The answer is positive. As shown previously, the strong kHz EM fluctuations behave as a temporal fractal following the persistent fBm model, while its temporal profile has a roughness in harmony with the universal spatial roughness of the fracture surfaces.

An important pursuit is to make a further quantitative comparison between characteristics of the kHz EM time series on one hand and the self asperity model on the other hand. Two relevant arguments are the following. 

{\it First argument}. The self asperity model (Hallgass et al., 1997) suggests that the distribution of areas of the asperities broken $A$ follows a power law 

\begin{displaymath}
P(A) \sim A^{-\delta},
\end{displaymath}

with an exponent $\delta$ which could be related to the Hurst exponent $0 < H < 1$ that controls the roughness of the fault. The former relation is obtained by supposing that the area of the broken asperities scales with its linear extension $l$ as $A_{asp} \sim 
l^{(1+H)}$. 

The $H$-exponent is distributed around the value 0.7. Based on the previously presented arguments it is reasonable to assume that in the case of the L'Aquila EQ the broken asperities scaled with its linear dimension $l$ as $A_{asp} \sim l^{1.7}$. Importantly, numerical studies performed by de Arcangelis et al., (1989) indicate that the number of bonds that break scales during the whole process of fracture as $l^{1.7}$ with the system size $l$. This consistency further supports the seismogenic origin of the anomalies under study

{\it Second argument}.  The self affine asperity model (Hallgass et al., 1997) also reproduces the Gutenberg-Richter law. More precisely, it suggests that a seismic event releases energy in the interval $[E, E+dE]$ with a probability $P(E)dE$, $P(E) \sim E^{-B}$, where $B = \alpha + 1$ and $\alpha = 1 - H/2$ with $\alpha \in [1/2,1]$. 

In the present case, the Hurst-exponent $H \sim 0.70$ leads to $\alpha \sim 0.65$. Thus, the fracture of asperities released EM energies following the distribution $P(E) \sim E^{-B}$, where $B \sim 1.65$. This value is in harmony with both geophysical and laboratory data. Indeed, the distribution of energies released at any EQ is described by the power-law, $P(E) \sim E^{-B}$, where $B \sim 1.4 - 1.6$ (Gutenberg and Richter, 1954). On the other hand, in laboratory scale, Houle and Sethna (1996) found that the crumpling of paper generates acoustic pulses with a power-law distribution in energy $P(E)=E^{-B}$, $B=1.3-1.6$. 

The observed consistencies further bridge the kHz EM anomalies under study with the activation of the L'Aquila fault.

\section {Similarities between the dynamics of EQs, kHz EM precursors, magnetic storms and epileptic seizures} 

A basic reason for our interest in ``complexity'' is the striking similarity in behavior near the global instability among systems that are otherwise quite different in nature. Empirical evidence has been mounting that supports the possibility that a number of systems arising in disciplines as diverse as physics, biology, engineering, and economics may have certain quantitative features that are intriguingly similar (Stanley, 1999, 2000; Sornette, 2002). 

Characteristically, de Arcangelis et al. (2006) presented evidence for universality in solar flare and EQ occurrence. Picoli et. all (2007) reported similarities between the dynamics of geomagnetic signals and heartbeat intervals. Kossobokov and Keilis-Borok (2000) have explored similarities of multiple fracturing on a neutron star and on the Earth, including power-law energy distributions, clustering, and the symptoms of transition to a major rupture. Sornette and Helmstetter (2002) have presented occurrence of finite-time singularities in epidemic models of rupture, EQs, and starquakes. Abe and Suzuki (2004) have shown that internet shares with 
EQs common scale-invariant features in its temporal behaviors. Peters et al., (2002) have shown that the rain events are analogous to a variety of nonequilibrium relaxation processes in nature such as 
EQs and avalanches.

Herein we focus on EQs, kHz EM precursors, magnetic storms, and epileptic seizures. Magnetic storms are the most prominent global phenomenon of geospace dynamics, interlinking the solar wind, magnetosphere, ionosphere, atmosphere and occasionally the Earth's surface (Daglis, 2001; Daglis et al., 2003). Magnetic storm intensity is usually represented by an average of the geomagnetic perturbations measured at four mid-latitude magnetic observatories, known as the $D_{st}$ index (http://swdcwww.kugi.kyoto-u.ac.jp/). Storm is an interval of time when a sufficiently intense and long-lasting interplanetary convection electric field leads, through a substantial energization in the magnetosphere-ionosphere system, to an intensified strong ring current enough to exceed some key threshold of the quantifying storm time $D_{st}$ index \cite. Recently, we showed that {\it the modified non-extensive Gutenberg-Richter type law (7) is also able to describe the distribution of magnitude of magnetic storms (Balasis and Eftaxias, 2009). We underline that the best-fit for this analysis is given by a $q$ parameter equal to 1.76}.

{\it It is very interesting to observe the similarity in the $q$-values associated with: (i) all the catalogues of EQs used in [(Solotongo and Posadas, 2004); (Silva et al., 2006)]. (ii) the precursory sequence of kHz EM bursts under study; and (iii) magnetic storms}.

Finally, electroencephalogram (EEG) time-series provide a window through which the dynamics of epileptic seizure preparation can be investigated under well-controlled conditions. Interestingly, theoretical studies (Hopfield, 1994; Herz and Hopfield 1995; Rundle et al., 1995; Corral et al., 1997) suggest that the EQ dynamics at the final stage and neural seizure dynamics should have many similar features and can be analyzed within similar mathematical frameworks. Authors have tested this universal hypothesis on the Earth's crust and the epileptic 
brain {\it verifying that epileptic seizures could be considered as quakes of the brain (Osorio et al., 2007; Eftaxias et al., 2006)}. Importantly, Osorio et al., (2007) have tested the concept of universality on the Earth's crust and the epileptic brain in terms of probability density functions of seismic moments and epileptic seizure energies. They showed that both statistics are compatible with the same power law with an exponent $\sim$ 2/3. We remind that sequence of kHz EM EQs under study follows a power-law type probability density function of 
magnitudes with an exponent $\sim$ 0.5, while that detected prior to the Athens EQ is characterized by an exponent 0.62 (Eftaxias et al., 2006 and references therein). 

{\it The above-mentioned results imply that a unified theory may exist for the ways in which firing neurons / opening cracks / sub-storms organize themselves to produce a significant epileptic seizure / a strong pre-earthquake EM burst or a significant EQ / intense magnetic storm}. The observed striking similarities: (i) Encourage a fruitful exchange of ideas and techniques among the above mentioned various different disciplines. (ii) Suggest a common approach to the interpretation of various phenomena in terms of the same driving physical mechanism. The universal character of the approach of various extreme phenomena is a challenge for our understanding of these phenomena (Stanley, 1999; 2000; Fukuda et al., 2003; Vicsek, 2001, 2002). Especially, increasing needs for prediction of geophysical and biological shocks have triggered rapid advances in the understanding of complexity.  

\section{Detection of ionospheric perturbations associated with the L'Aquila EQ} 

EQ precursory signatures appear not only in the lithosphere, but also in the atmosphere and ionosphere (Lithosphere-Atmosphere-Ionosphere Coupling) (e.g, Hayakawa and Fujinawa, 1994; Molchanov and Hayakawa, 1998; Hayakawa, 1999; Parrot 1999; Hayakawa and Molchanov, 2002; Pulinets et al., 2002; Muto et al., 2008; Uyeda et al., 2009). {\it Both natural signals (atmospherics) and artificial EM signals propagate in the Earth-ionosphere wavequide. Any change in the lower ionosphere may result in significant changes in the signal received at a station}. Especially, statistical analyses on the correlation between the lower ionospheric perturbations (as they detected by the ground-based reception of subionospheric EM waves from VLF/LF transmitters) have been performed by Rozhnoi et al (2004) and Maekawa et al (2006). The authors conclude that the lower ionosphere is definitely perturbed for the shallow EQs with magnitude larger than 6.0. Pulinets et al. (2003) have provided a strong evidence for occurrence of ionospheric precursors well before the main shock of EQ: ionospheric precursors within 5 days before the seismic shock were registered in 73

The L'Aquila EQ occurred in land was very shallow and its magnitude was 6.3. These characteristics justify the appearance of pre-earthquake ionospheric perturbations associated with the L'Aquila EQ. Indeed, EM anomalies possibly originated in an induced Lithosphere-Atmosphere-Ionosphere Coupling have yet reported. 

(i) Biagi et al. (2009), based on radio sounding, have observed such an anomaly. The intensity of MCO ($f = 216 kHz, France$, CZE ($f = 270 kHz, Czech Republic) and CLT ($f = 189 kHz, Sicily, Italy) broadcast signals have been collected by a receiver operating in a place located about 13 km far from the EQ. The LF signals are characterized by the ground-wave propagation and the sky-wave propagation. The sky-wave signal received by an antenna can be considered as a ray starting from the transmitter and reflected on or more times (hops) by the lower ionosphere and by the ground. {\it From 31 March to April 1 the intensity of the MCO signal dropped and this drop was observed only in this signals. More precisely, the anomaly is represented by the disappearance of the signal at day time and at night time}. Biagi et al. (2009) propose that at the time of the observed anomaly the ground wave practically is the day time signals and the sky wave (one hop) is the night time signal. 

(ii) Rozhnoi et al. (2009), based on very low frequency (kHz) radio sounding, have also reported that ionospheric perturbations appeared before the L'Aquila EQ. Using two known procedure of analysis: revelation of night-time signal anomalies and anomalous shift in the evening terminator time, they have found clear anomalies in the time interval 2--8 days before the EQ occurrence for seismic paths while the similar signal anomalies are absent for control paths.

A clear pre-seismic EM anomaly has been detected by the two Short Thin Wire Antennas of 100m length each, lying in the Earth's surface in the EW and NS directions, at the Zante station (see Fig. 1). This anomaly (see Fig. 5) was observed a few days before the L'Aquila EQ at Zante station. To be more concrete, the anomaly was continuously observed from March 29, 2009 up to April 2, 2009. Fig. 5 shows the chain of the daily ULF EM records. The record refers to the natural EM emission that depends mostly on sources rooted in atmosphere and ionosphere [see page 46 in ref. (Gokhberg et al. 1995)]. Fig. 5 stands up that the perturbation is represented by a significant decrease of the recorded ULF EM activity both at day time and night time. {\it We pay attention to the fact that our results show very close similarities in comparison to the anomalies observed by Biagi et al. (2009) and Rozhnoi et al. (2009) in terms of leading time}.

Clear similar anomalies have been also detected before significant (M > 6) shallow land-based EQs that occurred in Greece during the last year. These anomalies also emerged a few days before the EQ occurrence (e.g, Eftaxias et al., 2000, 2002, 2004; Karamanos et al., 2006).  

A reliable manifestation of lithosphere-ionosphere coupling during the preparations of the L'Aquila EQ may further prove the assumption that an important part of energy, and thus seismogenic EM emission, was transmitted from the lithosphere into the atmosphere and further into the ionosphere. 

\section {Conclusions}

Clear ULF, MHz and kHz EM anomalies were detected prior to the L'Aquila EQ that occurred on April 6, 2009. The detected anomalies followed this temporal scheme: (i) The MHz EM anomalies were detected on March 26, 2009 and April 2, 2009, respectively. (ii) The kHz EM anomalies emerged on April 4, 2009. (iii) The ULF EM anomaly was continuously recorded from March 29, 2009 up to April 2, 2009. 

``Are EQs predictable?'' is a question hotly debated in the EQ science community. Its answer begs for another question: "Are there credible EQ precursors?. Despite fairly abundant circumstantial evidence pre-earthquake EM signals have not been adequately accepted as real physical quantities. Therefore, the question effortlessly arises as to whether the observed anomalies were seismogenic or not. The main goal of this work is to provide some insight into this issue. 

The experimental arrangement affords us the possibility to determinine not only whether or not a single kHz, MHz, or ULF EM anomaly is pre-seismic in itself, but mainly whether a combination of such kHz, MHz, or ULF EM anomalies can be characterized as pre-seismic.

A major class of seismo-EM signals is rooted in anomalous propagation of EM signals over epicentral regions associated with the variation of Earth-ionosphere wave-guide due to a pre-seismic LAI-coupling. Importantly, Biagi et al., (2009) and Rozhnoi et al. (2009) have observed such as kHz EM anomalies in the time interval 2--8 days before the L'Aquila EQ. The recorded ULF EM anomaly at Zante station seems to fit in this category. Notice, the ULF EM anomaly shows similarity in comparison to the recorded kHz EM anomalies in terms of leading time. The seismo-ionospheric effects are rooted in {\it large-scale} anomalous phenomena appearing over the seismo-active area before {\it strong} EQs. The observed manifestation of LAI-coupling further proves the assumption that an important part of energy was transmitted from the lithosphere into the atmosphere and beyond during the preparation of the EQ under study. 

A second key class refers to EM signals believed to be emitted from within the focal zones (Uyeda et al., 2009). 
We argue that the detected MHz and kHz EM anomalies belong to this group. The performed analysis has been 
motivated by the following open questions}: (i) {\it How can we recognize a kHz or MHz EM observation as a 
pre-seismic one?. (ii) How can we link an individual EM precursor with a distinctive stage of EQ preparation 
process?. (iii) How can we identify crucial precursory symptoms in EM observations that indicate that the 
occurrence of the EQ is unavoidable?}. In our opinion the above questions are compelling questions for our 
understanding of pre-seismic EM emissions. Consequently, the main aims of this communication are threefold. 

First, we aim to suggest a procedure for the designation of the detected EM anomalies as seismogenic ones. We do not expect to be able to provide a succinct and solid definition of a pre-seismic EM emission. Instead, we aim, through a multidisciplinary analysis, to provide some elements of a definition. 

In Part I of this work (Eftaxias et al., 2009b) we focus on the detected kHz EM anomalies, which play a crucial role in our approach to the aforementioned three challenges. For this purpose, we analyze the data successively in terms of various concepts of entropy and information theory including, Shannon $n$-block entropy, conditional entropy, entropy of the source, Kolmogorov-Sinai entropy, $T$-entropy, approximate entropy, fractal spectral analysis, R/S analysis and detrended fluctuation analysis. Why use several tools? One cannot find an optimum organization or complexity measure. Thus, a combination of several such quantities which refer to different aspects is the most promising way. We argue that the performed multidisciplinary analysis clearly discriminates the recorded kHz EM anomalies from the background. They are characterized by: {\it (i) a significantly higher organization (or lower complexity, lower uncertainty, higher predictability, higher compressibility); and (ii) strong persistency}. The abrupt simultaneous appearance of both high organization and persistency implies that the underlying fracto-electromagnetic process is governed by a {\it positive feedback mechanism}, which expresses a positive circular causality that acts as a growth-generating phenomenon and therefore drives unstable patterns. That is to say, the property of {\it irreversibility} is included in the kHz EM activity under study. The field of study of complex systems holds that the dynamics of complex systems are founded on universal principles that may used to describe disparate problems ranging from particle physics to economies of societies. Importantly, the application of the procedure outlined above also extracts the {\it pathological symptoms of high organization and persistency} from other quite different complex catastrophic phenomena, such as epileptic seizures, magnetic storms and solar flares, as predicted by the theory of complex systems. Therefore, the simultaneous appearance of both high organization and persistency seems to constitute a universal feature of catastrophic events. Universal footprints guide and simplify our inquiries into the study of specifics. 

However, by the end of Part I (Eftaxias et al., 2009b) we conclude that the extracted crucial signatures from the kHz EM anomalies by themselves are not sufficient to characterize the anomalies as pre-seismic. In our opinion the symptoms of high organization and persistency constitute {\it necessary} but not {\it sufficient} criteria in a succinct and solid definition of a pre-seismic kHz EM anomaly. They provide some insight into how to designate pre-earthquake kHz EM emissions and identify and recognize some features beyond which the evolution toward global failure becomes unavoidable. However, they cannot link the kHz EM anomaly with a characteristic stage of the EQ preparation process. Therefore, in the present Part II of this work we aim to link the detected MHz and kHz EM anomalies with equivalent last stages of the EQ preparation process.

We paid attention to the fact that the MHz radiation appears earlier than the kHz on both laboratory and geophysical scales (Eftaxias et al., 2002). Seismologists have never directly observed rupture in Earth's interior. Instead, they glean information from seismic waves. In this logic, physicists attempt to link the available different EM observations to the consecutive processes occurring in the Earth's interior to help them interpret the data. We recently proposed the following two-stage model (Kapiris et al., 2004; Contoyiannis et al., 2005, 2008; Papadimitriou et al., 2008; Eftaxias et al., 2006, 2007b): (i) The first epoch, which includes the initial emergent MHz EM emission, is thought to be due to the fracture of a highly heterogeneous system that surrounds a family of large high-strength asperities distributed along the fault sustaining the system. (ii) The second epoch, which includes the emergent strong impulsive kHz EM radiation, is due to the fracture of the asperities themselves. The relevant experience gives the possibility to compare the results of the present study with previous ones. We argue that the sequence of MHz and kHz EM anomalies under study follows the aforementioned two stages model. 

First, we focus on the recorded MHz EM anomaly. Fracture process in heterogeneous materials is characterized by fundamental properties which should be mirrored on the precursor. The interplay between the heterogeneities and the stress field in the focal area should be responsible for the appearance of $\it anti-persistency$ in a seismogenic MHz EM time series. Furthermore, in analogy to the study of critical phase transitions in statistical physics, it has been proposed that the fracture of heterogeneous materials could be described {\it in analogy with a thermal continuous second order phase transition}. These features introduce two {\it necessary} criteria in the definition of a seismogenic MHz EM anomaly. (Herrmann and Roux, 1990; Sornette, 2004). We showed that the detected precursory MHz anomalies show antipersistency and can be described by means of a thermal second order phase transition.  

A fundamental question that we ought to address in the study of EM precursors is as follows. ``Is the evolution towards global failure irreversible after the appearance of the critical anti-persistent MHz EM precursor?'' We clearly state  that the detection of such as MHz EM anomaly does not mean that the occurrence of EQ is unavoidable. Its appearance reveals that the fracture of heterogeneous system in the focal area has been obstructed along the backbone of asperities that sustain the system: {\it the ``siege'' of strong asperities begins}. {\it The EQ will occur if and when the local stress exceeds the fracture stress of asperities}. We suggest that the abrupt emergence of kHz EM emissions reveals the fracture of asperities and thus inform us that the evolution of the process toward global failure is irreversible. 

We further investigate the seismogenic origin of the detected kHz EM anomaly. Despite the complexity of the Earth's crust, there are several universally holding scaling relations. Fundamental structural patterns of fracture and faulting process should be reflected on a kHz EM activity. Therefore, an important pursuit is to make a quantitative comparison between {\it temporal fractal patterns} included in the candidate precursory kHz EM time series on one hand, and universal {\it spatial fractal patterns} of fracture surfaces on the other hand.

Fracture surfaces were found to be self-affine following the fBm model over a wide range of length scales. 
Additionally, the spatial roughness of fracture surfaces has been interpreted as a universal indicator of surface 
fracture, weakly dependent on the nature of the material and on the failure mode. Therefore, two questions arise 
as to whether the kHz EM activity: (i) constitutes a temporal fractal following the fBm model; and (ii) has a 
roughness, which is in harmony with the universal spatial roughness of fracture surfaces. We showed that both two 
questions have positive answer: {\it the sequence of kHz EM pulses observed prior to the L'Aquila EQ could be 
induced by the slipping of two rough and rigid persistent fBm profiles one over the other.}

Consequently, the self-affine nature of fracture surfaces and the universal character of their roughness insert two {\it necessary and sufficient} criteria in the definition of a kHz fracto-EM emission. These two criteria identify and recognize crucial features beyond which the evolution toward global failure becomes unavoidable.   

The aspects of self-affine nature of faulting and fracture have been also well-established. Consequently, the questions that are asked are whether the kHz EM activity detected prior to the L'Aquila EQ behaves as a: (i) reduced image of the regional seismicity; and (ii) magnified image of laboratory seismicity. We showed that the above two questions also have positive answer: the sequence of kHz EM pulses observed prior to the L'Aquila EQ could be induced during the fracture of the backbone of strong and large entities that where distributed along the activated fault sustaining the system.

Therefore, the self-affine nature of faulting and fracture add two more {\it necessary and sufficient} criteria in the definition of a kHz fracto-EM emission. These two new criteria also identify and recognize crucial features beyond which the evolution toward global failure becomes unavoidable. 

As it was repeatedly mentioned we do not expect to be able to provide a succinct and solid definition of a pre-seismic MHz and kHz sequence of EM emission. However, we think that through a multidisciplinary analysis we provide some elements of such a definition. The combination of the introduced criteria with the proposed two stages model of EQ preparation process in terms of MHz and kHz EM emissions increases the efficiency of the proposed scheme and probably opens up new possibilities for prediction of EQs. It is attractive that the suggested criteria include only universal parameters and do not involve physical and mechanical parameters of materials which are beyond the control of the researcher. 

The additional appearance of any other different pre-seismic indicator, further verifies the seismogenic origin of each single link of the observed candidate precursory chain. The observed ULF EM anomaly a few days before the L'Aquila EQ, which is probably due to a pre-seismic LAI-coupling, strongly supports the seismogenic origin of the detected MHz-kHz EM anomalies. 

To conclude, the comprehensive understanding of the EM response to geodynamic processes occurring in the upper 
crust of the Earth is a challenging task in modern geophysics. Whether EM precursors to EQs exist is an important 
question not only for EQ prediction but also for understanding the physical processes of EQ generation (Uyeda et 
al., 2000). In physics, the degree to which we can predict a phenomenon is often measured by how well we 
understand it. In this communication we have tried to provide some insight into how to designate the pre-seismic 
kHz-MHz EM emissions in order to better understand the preparation process of EQs in them. While the 
results described here seem to be tolerable, only a small fraction of the wealth of unexplored possibilities has 
been made use of. Whether the presented ideas will prove to be universal or disappear as others, will turn out in 
the future. The complexity of pre-seismic EM activities generation is enormous, and thus a huge amount of 
research is needed before we begin to understand it. It is a certainty that the problems of societal response to 
EQ predictions are not going to be solved until scientific problems can be brought under control. These are no 
more difficult than they were several decades ago; they are only more clearly defined today. We recognize today 
that the scientific problems are not simple. In any case, scientists must initiate shifting the minds of 
community from pessimistic disbelieve to optimistic challenging issues of EQ predictability, based on the recent 
enormous progress in real-time retrieval and monitoring of distributed multitude of a multiplicity of data.

\newpage

\begin{figure}[t]
\begin{center}
\end{center}
\caption{In the map the location of The Zante station and the epicentre of the L'Aquila EQ are shown.}
\end{figure}

\begin{figure}[t]
\begin{center}
\includegraphics[width=8.3cm]{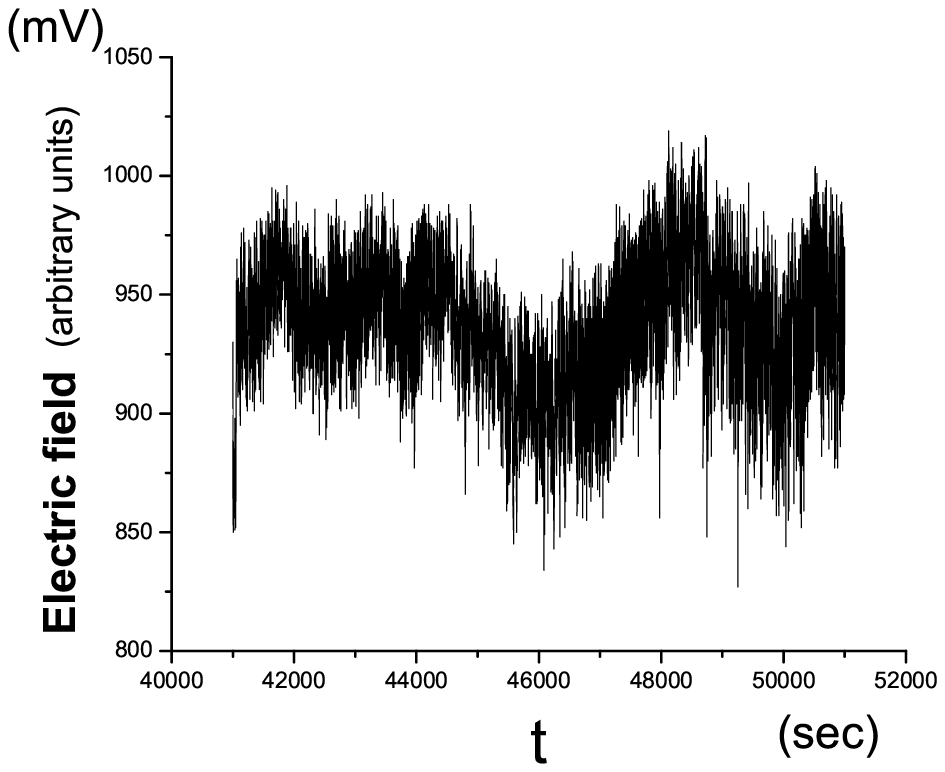}\\
\includegraphics[width=8.3cm]{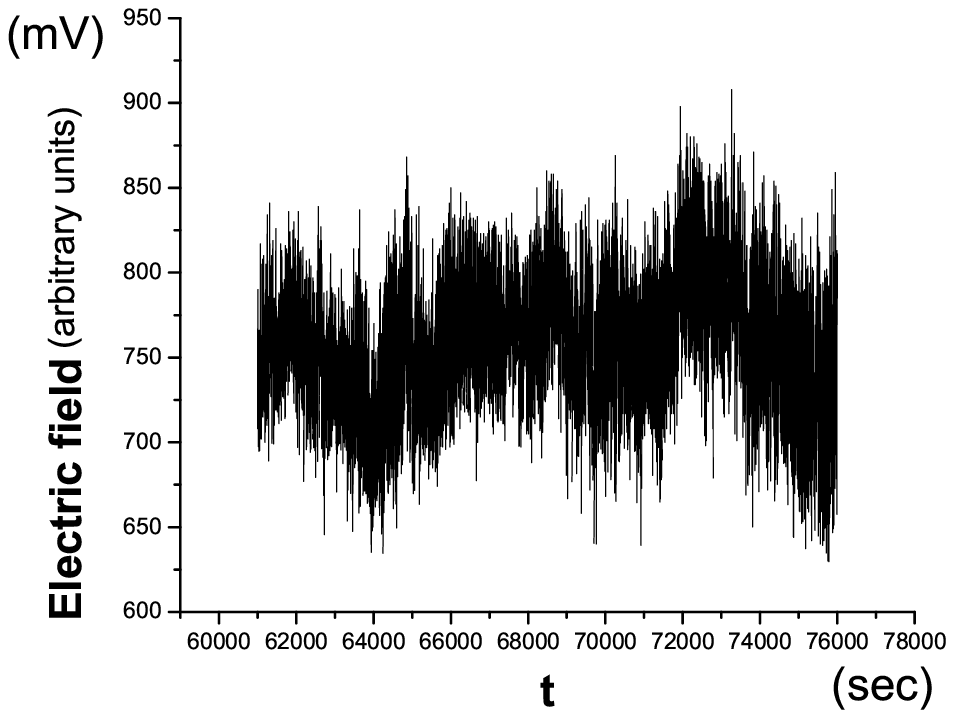}
\end{center}
\caption{The excerpts of the 41 MHz magnetic field strength time series which show anti-persistency and behave as a continuous (second order) thermal phase transition. The two time intervals, which include critical fluctuations, emerged on March 26 2009 (upper panel) and April 2 2009 (lower panel), respectively.}
\end{figure}

\begin{figure}[t]
\begin{center}
\includegraphics[width=8.3cm]{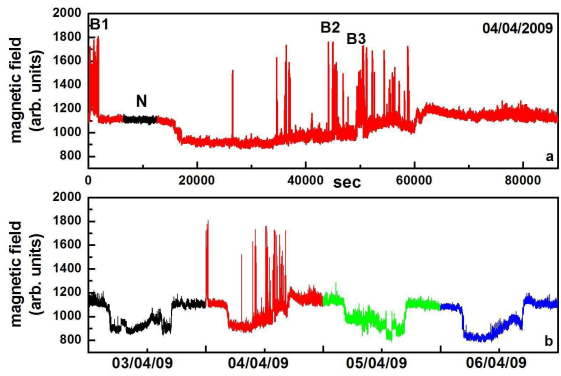}
\end{center}
\caption{(a) A sequence of strong EM impulsive bursts at 10 kHz emerged on April 4 2009. (b) These anomalies appeared over a quiescence period concerning the detection of EM disturbances. A segment from the EM background (N) and three excerpts of the detected strong kHz EM activity (B1, B2, B3) 
have been marked in the time series of April 4 2009.}
\end{figure}

\begin{figure}[t]
\begin{center}
\includegraphics[width=8.3cm]{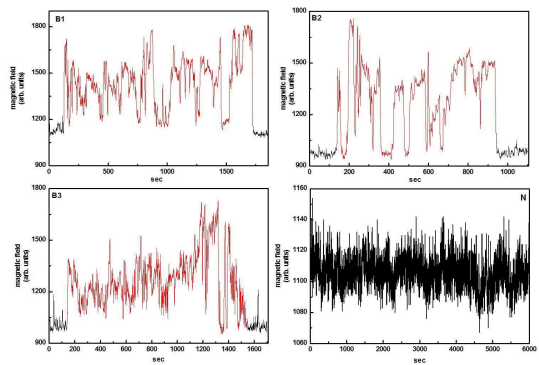}
\end{center}
\caption{Magnified images of the excerpts N, B1, B2, and B3 that are shown in Fig. 2.}
\end{figure}

\begin{figure}[t]
\begin{center}
\includegraphics[width=8.3cm]{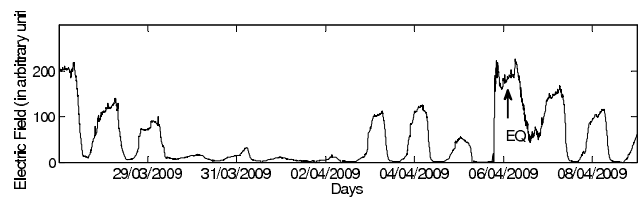}
\end{center}
\caption{The ULF EM anomaly which was detected from March 29, 2009 up to April 2, 2009.} 
\end{figure}

\begin{figure}[t]
\begin{center}
\includegraphics[width=8.3cm]{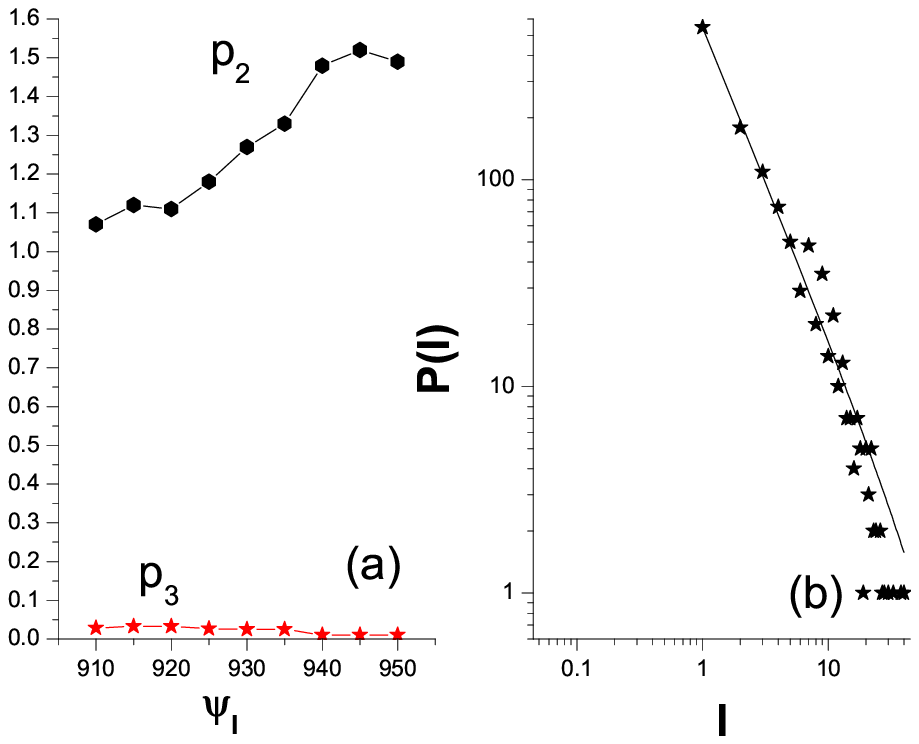}
\end{center}
\caption{The figure refers to the critical window which appeared on March 26, 2009 [see Fig. (2)]. The Fig. (6a) shows the exponents $p_2$ and $p_3$ versus the end of laminar region $\psi_l$. We observe that the majority of 
the $p_2$ values is greater than 1 and the corresponding $p_3$-values are close to zero for all the end points of laminar regions $\psi_l$.  This finding suggests that the laminar lengths tend to follow a power-law type distribution. The distribution of the laminar lengths for a given end point $\psi_l$ is shown in Fig. (6b). The function $P(l) \sim  l^{-p_2} e^{-p_3 l}$ fit the data with $p_2 = 1.48$ and $p_3 = 0.009$, while the quality of fitting is excellent ($r^2 = 0.998$).}
\end{figure}

\begin{figure}[t]
\begin{center}
\includegraphics[width=8.3cm]{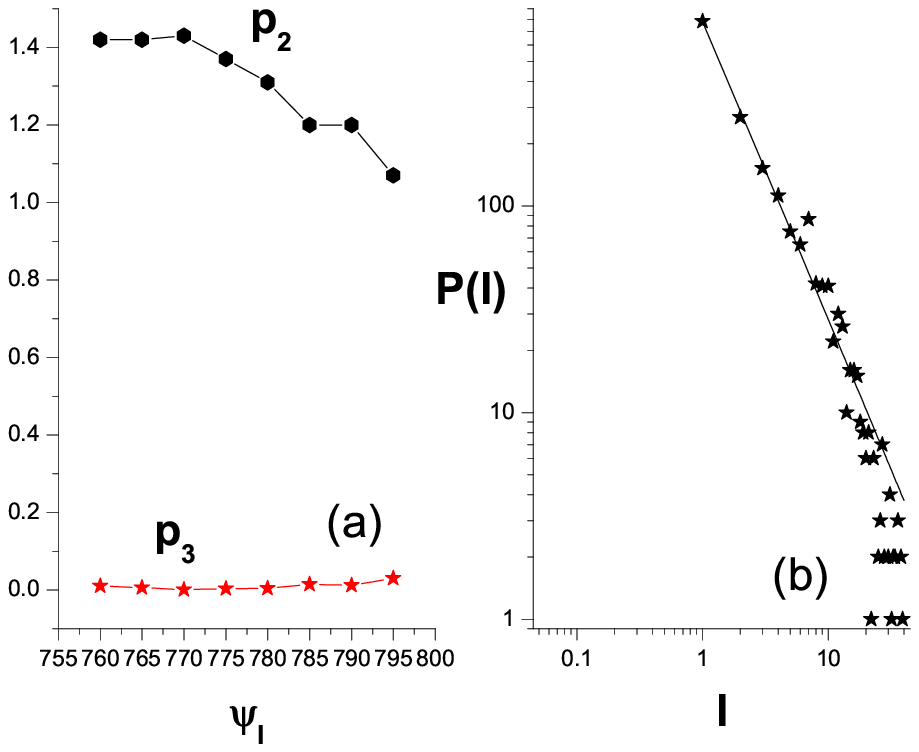}
\end{center}
\caption{The figure refers to the critical window which appeared on April 2, 2009 [(see Fig. (1)]. The Fig. (7a) shows the exponents $p_2$ and $p_3$ versus the end of laminar region $\psi_l$. We observe that the majority of the 
$p_2$ values is greater than 1 and the corresponding $p_3$-values are close to zero for all the end points of laminar regions $\psi_l$. This finding suggests that the laminar lengths tend to follow a power-law type distribution. The distribution of the laminar lengths for a given end point $\psi_l$ is shown in Fig. (7b). The function $P(l) \sim  l^{-p_2} e^{-p_3 l}$ fit the data with $p_2 = 1.43$ and $p_3 = 0.008$, while the quality of fitting is excellent ($r^2 = 0.997$).}
\end{figure}

\begin{figure}[t]
\begin{center}
\includegraphics[width=8.3cm]{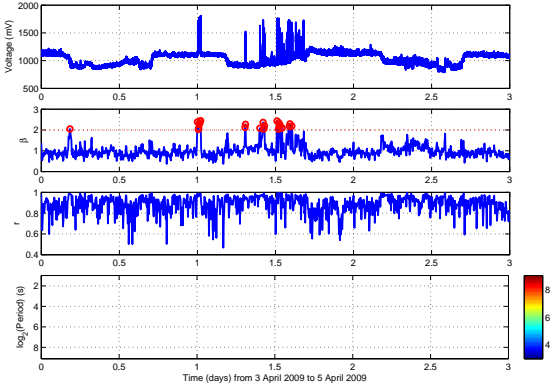}
\end{center}
\caption{From top to bottom are shown the 10 kHz time series, spectral exponents $\beta$, linear correlation coefficients $r$, and the wavelet power spectrum from April 3, 2009 to April 5, 2009. The red dashed line in the plot marks the transition from anti-persistent to persistent behavior. We observe that in the emerged strong kHz EM bursts the coefficient r takes values very close to 1, i.e., the fit to the power-law $S(f) \propto f^{-\beta}$ is excellent. This means that the fractal character of the kHz EM activity is solid. The $\beta$ exponent takes high values, i.e., between 2 and 3. This reveals that the candidate precursor follows the persistent Fractional Brownian motion model.} 
\end{figure}

\begin{figure}[t]
\begin{center}
\includegraphics[width=8.3cm]{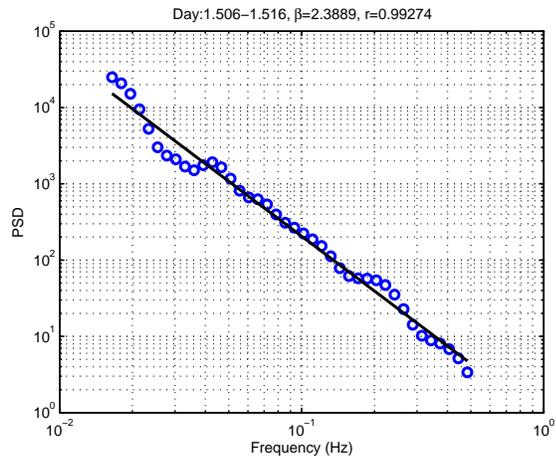}
\end{center}
\caption{The $log$-$log$ representation of the function $S(f) \propto f^{-\beta}$ into an excerpt of the recorded kHz EM anomaly, namely EM burst B1 [see Figs. (3), (3)], is shown. The associated $\beta$ exponent is equal to 2.39, 
which leads to $H = 0.7$. This means that the roughness of the profile of the kHz EM timeseries is in harmony with the universal roughness of fracture surfaces.}
\end{figure}

\begin{figure}[t]
\begin{center}
\includegraphics[width=8.3cm]{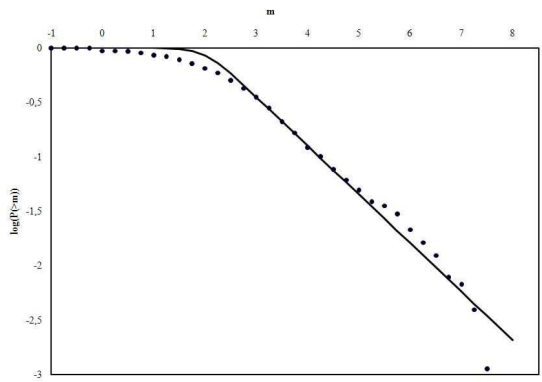}
\end{center}
\caption{We use formula (7), which quantitatively describes the nonextensive model for EQ dynamics in terms of regional seismicity, to calculate the relative cumulative number of kHz ``electromagnetic earthquakes'', $P(>m)$ ,  included in the whole precursory phenomenon depicted in Fig. 3. There is an agreement of formula (7) with the data. The associated nonextensive $q$-parameter has value 1.82.}
\end{figure}

\begin{figure}[t]
\begin{center}
\includegraphics[width=8.3cm]{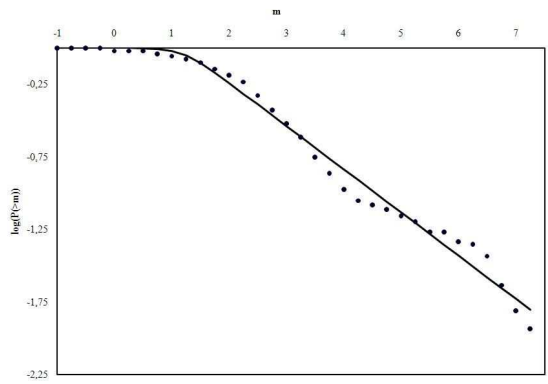}
\end{center}
\caption{We use formula (7), which quantitatively describes the nonextensive model for EQ dynamics in terms of regional seismicity, to calculate the relative cumulative number of kHz ``electromagnetic earthquakes'', $P(>m)$ ,  included in the first burst (B1) depicted in Figs. (2), (3). There is an agreement of formula (7) with the data. The associated nonextensive $q$-parameter has a value of 1.87.}
\end{figure}

\begin{figure}[t]
\begin{center}
\includegraphics[width=8.3cm]{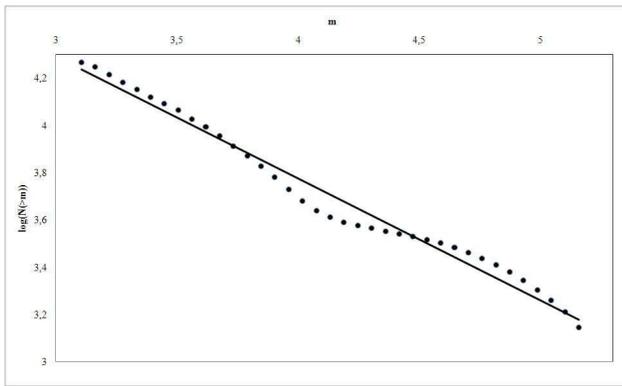}
\end{center}
\caption{Number of kHz ``electromagnetic fluctuations'' emerged on April 4, 2009 with a magnitude $m$ higher than that given by the corresponding abscissa. The continuous line is the least squares fit the law $\log N(>m)=\alpha-bm$, where $b \sim 0.52$.}
\end{figure}

\end{document}